# Application of Hilbert-Huang decomposition to reduce noise and characterize for NMR FID signal of proton precession magnetometer

Liu Huan[1,2,3,4], Dong Haobin[2, *], Zheng Liu[2], Jian Ge[1], Bingjie Bai[3], Cheng Zhang[1]

(1．School of Automation, China University of Geosciences, Hubei Wuhan 430074, China；

2．School of Engineering, University of British Columbia Okanagan Campus, BC, Kelowna V1X 1V7, Canada；

3．Institute of Geophysics & Geomatics, China University of Geosciences, Hubei Wuhan 430074, China；

4. Hubei Key Laboratory of Advance Control and Intelligent Automation for Complex Systems, Hubei Wuhan 430074, China)

**Abstract**：The parameters in a nuclear magnetic resonance (NMR) free induction decay (FID) signal contain information that is useful in magnetic field measurement, magnetic resonance sounding (MRS) and other related applications. A real time sampled FID signal is well modeled as a finite mixture of exponential sequences plus noise. We propose to use the Hilbert-Huang Transform (HHT) for noise reduction and characterization, where the generalized Hilbert-Huang represents a way to decompose a signal into so-called intrinsic mode function (IMF) along with a trend, and obtain instantaneous frequency data. Moreover, the HHT for an FID signal's feature analysis is applied for the first time. First, acquiring the actual untuned FID signal by a developed prototype of proton magnetometer, and then the empirical mode decomposition (EMD) is performed to decompose the noise and original FID. Finally, the HHT is applied to the obtained IMFs to extract the Hilbert energy spectrum, to indicate the energy distribution of the signal on the frequency axis. By theory analysis and the testing of an actual FID signal, the results show that, compared with general noise reduction methods such as auto correlation and singular value decomposition (SVD), combined with the proposed method can further suppress the interfered signals effectively, and can obtain different components of FID signal, which can use to identify the magnetic anomaly, the existence of groundwater etc. This is a very important property since it can be exploited to separate the FID signal from noise and to estimate exponential sequence parameters of FID signal.

**Key Words**：Free induction decay signal；Hilbert Huang transform；Empirical mode decomposition；Noise reduction；Nuclear magnetic resonance

# I. Introduction

Nuclear magnetic resonance (NMR) free induction decay (FID) signals, useful in weak magnetic measurement [1, 2], allow for the noninvasive detection and characterization of groundwater [3, 4] etc., are modeled as a finite mixture of exponential functions or sequences. The estimation of parameters in the model, including amplitudes, phases and frequencies, becomes an important issue in magnetic field measurement and evaluation of related applications. If there is no noise in the FID signals, the SNR will be higher, thereby improving the measurement accuracy and resolution [5]. In addition, the effective identification of groundwater existence can also be enhanced. Unfortunately, most real FID signals have relatively very high noise [6]; noise reduction, therefore, becomes an important, yet difficult problem in NMR FID signal analysis. Moreover, since the exponential decay characteristic of FID signal, which will decay rapidly to zero in no more than one second, especially when there exists magnetic field anomaly; one should face the initial parameter selection problem and useless component separation problem that are sensitive to the measurement and analysis procedure.

Various methods for noise reduction of NMR FID signals have been proposed. The concepts of correlation adaptive filtering [7] and signal processing techniques by Gabor expansion [8] and Prony [9] are stable, mature and well developed. Wang proposed a matrix mathematical model constructed by FID signal [10], by combining with frequency division multiplexing (FDM), the results prove that the matrix mathematical model construction based on FDM can meet the requirement of practical application. However, these methods stayed only in the stage of simulation, the performance need to be further verified. There are some methods that have been applied to practical scenarios. A FID signal processing technique based on singular value decomposition (SVD) & short-time Fourier transform (STFT) was proposed in [11], the simulation and practical results indicate that this method can suppress the noise efficiently, as well as the improvement of environmental adaptability. Denisov proposed a linear regression method, it can not only widen the registration bandwidth, but also relax the requirements to the hardware [12]. In literature [13], the typical different noises of FID signal, such as narrow band noise, phase noise and circuit noise were further investigated and analyzed. In addition, many other FID signal processing methods also have been proposed [14-17], and the results are encouragement. The key issue is how to reduce the noise either directly or indirectly to improve the measurement performance. However, most of these methods show good results only for high SNR, and ignore the case of noise reduction based on signal decomposition.

In this paper, we adopted a method called Hilbert-Huang transform (HHT) to process the FID signal. The generalized HHT represents a time-sampled signal in terms of a collection of time-shifted and frequency-distributed versions of a single sequence (FID sequence). Aimed to further identify the performance of the proposed method, we developed a prototype of proton precession magnetometer. With a real collected FID sequence, a new component model of FID signal is presented by combining the HHT. Intrinsic mode functions (IMFs) in the high frequency or low frequency parts, which are the trend of original pure FID signal data, are obtained to replace the traditional motive average lines to analyze the data. The practical test results show that, by combining the proposed signal processing method with some commonly used filtering algorithms, the performance of noise reduction give differences better than those of traditional methods. Moreover, the proposed method can be used to split the signal into original and noise, eliminate the noise component, thereby further suppressing the noise.

## II. FID signal

The amplitude of the FID signal generated by Larmor precession effect can be written as [18]:

$$\varepsilon(t) = CM\omega e^{t/T_2} \sin\omega t$$

where, $C$ stands for a constant value, $M$ is the proton's magnetic intensity, $\omega$ is the angular frequency, and $T_2$ is the transverse relaxation time, which is usually less than one second.

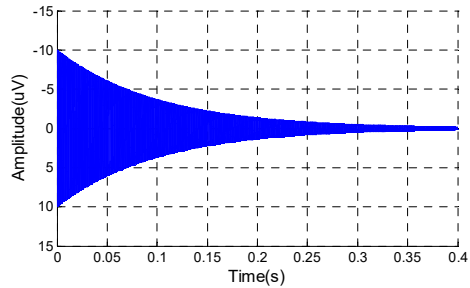

Fig. 1. The schematic diagram of FID signal.

Figure 1 shows the waveform, we observe that the FID signal is a sine wave with an exponential decay of the magnitude. In addition, the FID is a signal at micro volt level with considerable noises, which lead to the difficult to measure.

## III. Hilbert-Huang Transform

Hilbert-Huang transform (HHT) is a new method for analyzing nonlinear and non-stationary signals which is proposed by Huang, et al. in 1998 [19]. Compared with wavelet method, HHT can eliminate the spurious harmonics that have no physical mean, and is superior to wavelet method in some applications [20]. The HHT has shown to be useful for decomposing sunspot number into its intrinsic frequency components. Also, one of the main strengths of the HHT method is to compare the frequency components of two or more signals to determine relationships between them. One of the crucial part of this method is the Empirical Mode Decomposition (EMD), which can decompose the complex signals into a finite and small number of Intrinsic Mode Functions (IMF) from high frequency to lower frequency [21]. Then the different frequency oscillations for each signal can be compared directly and checked for correlations. Generally, the decomposed signals are adaptive and highly efficient, and the last decomposed signal is the trend. Another crucial part is Hilbert transform, which is used to the decomposed IMFs and construct the energy-frequency-time distribution, designated as the Hilbert spectrum, from which the time localities of events will be preserved.

*A. Empirical Mode Decomposition*

Empirical mode decomposition (EMD) is the first stage in HHT, which decomposes the signal into a set of IMFs by a series of shifting's. The IMF is stationary derived from EMD with shifting process, it should satisfy two conditions:

1) The number of extrema and zero crossings in a given data set should either equal or differ at most by

one;

2) The mean value of the envelop defined by the local maxima and minima should be equal to zero. Hence, the shifting process are as following for a given signal $f(t)$:

  a) Find out local maxima $f(t)_{max}$ and minima $f(t)_{min}$ of the signal;
  b) trace the envelope of maxima and minima with suitable spline;
  c) find out the mean value of maxima and minima at each point;

$$m(t) = \frac{f(t)_{max} + f(t)_{min}}{2} \tag{3.1}$$

  d) subtract the mean value $m(t)$ from the original signal $f(t)$;

$$e(t) = f(t) - m(t) \tag{3.2}$$

  e) return to step a), stop when $e(t)$ remains nearly unchanged;
  f) Once we obtained an IMF function $u(t)$, remove it from the signal:

$$f(t) = f(t) - u(t) \tag{3.3}$$

and return to step a) if $f(t)$ has more than one extreme, generally neither a constant or a trend.

### B. Hilbert Transform

Having obtained the intrinsic mode function components, there will have no difficulties in applying the Hilbert transform to each component. Hilbert transform is the convolution of signal $f(t)$ with $1/t$ which preserves the time format of the signal's amplitude and frequency [22]. Spectral estimation is the second step of the HHT. This consists in computing the instantaneous frequency for each IMF using the Hilbert Transform and analytic signal concept. The Hilbert Transform of $f(t)$ is given as:

$$H(t) = \frac{1}{\pi} P \int \frac{f(t')}{t - t'} dt' \tag{3.4}$$

where, $P$ is the Cauchy principle value. Analytic signal $A(t)$ combines with $H(t)$ and $f(t)$. $f(t)$ is the real part of $A(t)$ while $H(t)$ is the imaginary part. The $A(t)$ can be written as:

$$A(t) = f(t) + iH(t) = a(t)e^{i\theta(t)} \tag{3.5}$$

the amplitude and phase of $A(t)$ are:

$$\begin{cases} a(t) = \sqrt{f^2(t) + H^2(t)} \\ \theta(t) = arctg(\frac{f(t)}{H(t)}) \end{cases} \tag{3.6}$$

and the definition of instantaneous frequency is:

$$\omega(t) = \frac{d\theta(t)}{d(t)} \tag{3.7}$$

Performing the Hilbert transform for each IMF component, the original signal can be represented as the real part (*Re*) in the following [19]:

$$f(t) = \mathrm{Re} \sum_{n}^{i=1} a_i(t) \exp(i \int \omega_j(t) dt) \qquad (3.8)$$

where, $j = (-1)^{1/2}$. Therefore, the original signal can be represented as:

$$H(\omega, t) = \sum_{n}^{i=1} a_i(t) \exp(i \int \omega_j(t) dt) \qquad (3.9)$$

## IV. Experimental test and results

In practical environment, except for white noise, the FID signal is also accompanied by interference with phase noise, colored noise, singular noise and random noise [23, 24]. To further verify the effectiveness of HHT processed method for FID signal, a prototype system of proton magnetometer was developed as shown in figure 2, which is divided into two parts: the sensor and the signal conditioning system. The structure of the sensor is a dual-coil differential receiver, which can enhance the common mode rejection ratio of the circuit and effectively suppress the external interference [25]. The signal conditioning system consists of the tuning capacitor, amplifier, narrow band pass filter and comparator.

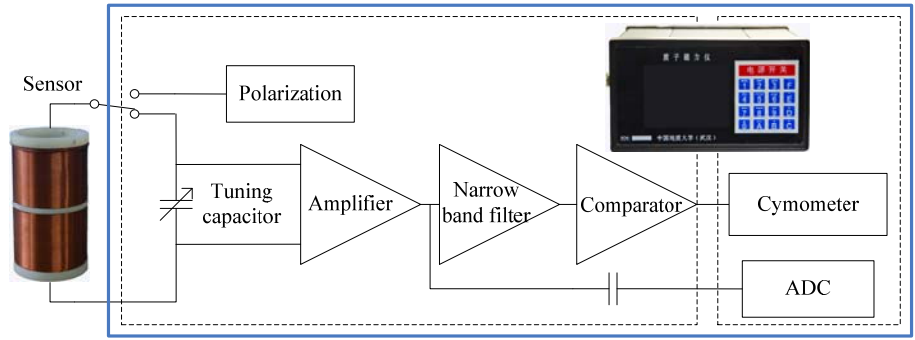

Fig. 2. Schematic diagram of the prototype.

The magnetic field of the test location with less interference is approximately 49323nT, corresponding to a frequency of 2100Hz. During the test, the FID signal output of the sensor was amplified and filtered, and then the analog to digital converter (ADC) was applied to collect the data. The sampling point is 2048 and the sampling rate is 20kHz, so the frequency resolution is 9.77Hz and the sampling period is about 102.4ms. In addition, a commercial proton magnetometer was adopted to verify the effectiveness of measurement results in the test location. The distance between two sensors is about 3 meters so that two instruments cannot be disturbed by each other.

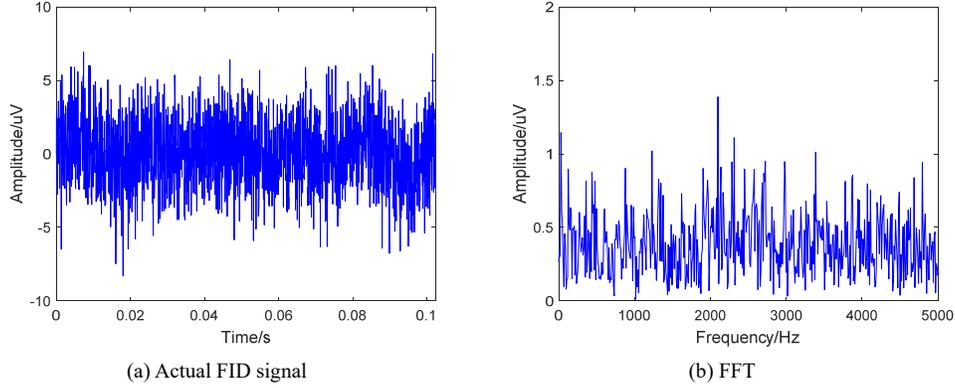

(a) Actual FID signal  (b) FFT

Fig. 3. The practical FID signal collected by ADC and FFT.

To further obtain the original feature of FID signal, the tuning system was not adopted. Figure 3 shows the collected actual untuned FID signal and the FFT results; according to the sampling point and sampling rate, the sampling time is 102.4ms. From figure 3 we observe that the practical signal has almost been drowned by noise, and the exponential decay trend of the signal cannot be observed clearly. From the FFT result of the original signal, the center frequency is around 2100Hz, corresponding to the FID signal frequency of the magnetic field at that test location. The interference of the noise with different frequency is serious, especially around 1000Hz and 4000Hz, which related to the test location, electromagnetic and electrical characteristics [11]. However, only based on the FFT, we cannot further analysis the detailed spectral characteristics of FID signal. Then the next step, we adopted the HHT to process FID signal.

### A. The Sifting Process by Empirical Mode Decomposition

As theoretical described above, the process of FID signal is indeed like sifting: to separate the finest local mode from the data first based only on the characteristic time scale. The sifting process, however, has two effects: (a) to eliminate original FID pure signal; (b) to smooth uneven amplitudes. While the first condition is absolutely necessary for the instantaneous frequency to be meaningful, the second condition is also necessary in case the neighboring wave amplitudes have too large a disparity. Unfortunately, the second effect, when carried to the extreme, could obliterate the physically meaningful amplitude fluctuations. Therefore, the sifting process should be applied with care, for carrying the process to an extreme could make the resulting IMF a pure frequency modulated signal of constant amplitude. To guarantee that the IMF components retain enough physical sense of both amplitude and frequency modulations, we have to determine a criterion for the sifting process to stop. This can be accomplished by limiting the size of the standard deviation ($S$), which is computed from the two-consecutive sifting. A typical value for $S$ can be set between 0.2 and 0.3 [19]. Overall, IMF1 should contain the finest scale or the shortest period component of the signal. We can separate IMF1 from the rest of the data by $s(t)$ - IMF1 = $R_1$. The $R_1$ called residue, since it contains information of longer period components, it is treated as the new data and subjected to the same sifting process as described above. This procedure can be repeated on all the subsequent, and the result is $R_1$ - IMF2 = $R_2$, ... , $R_{n-1}$ - IMFn = $R_n$. The sifting process can be stopped by any of the following predetermined criteria: either when the component IMFn or the residue $R_n$, becomes so small that it is less than the predetermined value of substantial consequence, or when the residue $R_n$ becomes a monotonic function from which no more IMF can be extracted. Even for data with zero mean, the final residue can still be different from zero; for data with a trend, then the final residue should be that trend. Thus, we achieved a decomposition of the data into n-empirical modes, and a residue $R_n$, which can be either the mean trend or a constant. To apply the empirical mode decomposition (EMD) method, a mean or zero reference is not required; EMD only needs the locations of the local extrema. The zero references for each component will be generated by the sifting process.

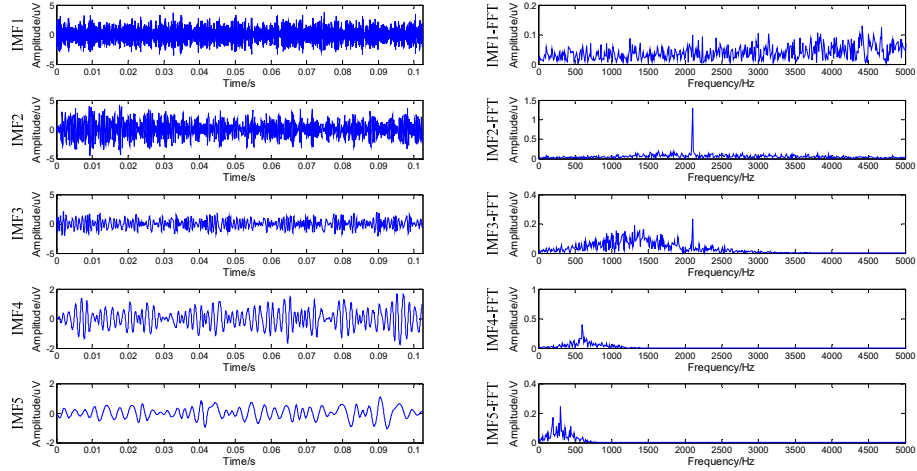

(a) The components IMF1-IMF5 and corresponding FFT

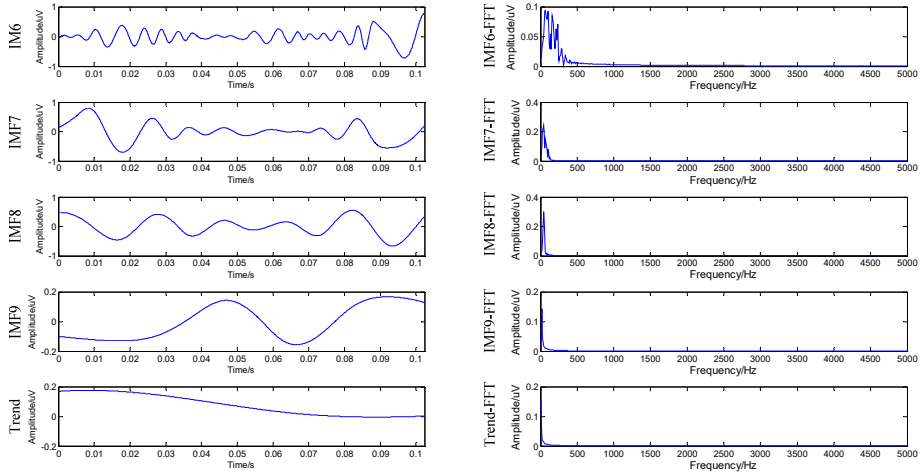

(b) the components IMF6-IMF9 and the trend, and corresponding FFT

Fig. 4. The results EMD comonents from the original FID signal data.

To illustrate the sifting process, we use the set of FID signal data mentioned above, the measured FID signal was processed based on the steps of HHT, first, the signal was decomposed by empirical mode, after decomposition, IMFs were analyzed. In order to make sure the effectiveness of this test, we have collected the untuned FID signal multiple times, and the number of IMFs is determined by the environmental background, especially when there exist interference around the sensor, such as ferromagnetic materials. Due to the environment of test location is unchanged, the number of IMFs is always constant as shown in figure 4. We observe that the number of IMFs is 10, and notice that the last component is not an IMF, it is the trend of the FID signal. From IMF1, we realize that there exist a lot of white noise in FID signal, which is distributed and having almost equal intensity at different frequencies, as well as a constant power spectral density; from IMF2, the center frequency is about 2100Hz, which is consistent with the signal frequency of the test location, and this would be regarded as the main useful FID signal. Comparing with the original signal in figure 3(a), the SNR has been improved effectively and the amplitude-frequency characteristic is more clearly; from IMF3, there is also a maximum amplitude value at 2100Hz, while the relatively large unknown noise are distributed in the frequency range from 500Hz to 2000Hz; from IMF4, there is a maximum amplitude value around 600Hz, while from IMF5 and IMF6, the noise are mainly distributed within 500Hz; from IMF7, IMF8 and IMF9, the noise are almost belong to low frequency signal, especially the power frequency interference of 50Hz; the last one of IMFs is the trend of signal, we can see that the longer the time, the smaller the amplitude of the signal, which is consistent with the exponential decay characteristic of FID signal. Comparing these with the traditional fourier expansion, one can immediately see the efficiency of the EMD: the expansion of a untuned FID data set with only ten terms. From the results, one can see a general separation of the data into locally non-overlapping time scale components. In some components, such as IMF2 and IMF4, the signals are intermittent, then the neighboring components might contain oscillations of the same scale, but signals of the same time scale would never occur at the same locations in two different IMF components.

There are several factors that may lead to the untuned FID signal contains so much noise signal, such as the effect of power frequency interference (50Hz / 60Hz) in natural environment, the audio signal, the magnetic field anomaly signal around the test location etc. The noise mentioned above belong to the external noise, and the internal noise is also the factor that should not be ignored, such as the electrical properties of electronic device, further information about the electrical noise research can be found in [11].

### B. Completeness and Orthogonality

By virtue of the decomposition, completeness is given. As a check of the completeness for the FID data numerically, we can reconstruct the data from the IMF components starting from the longest to the shortest periods in the sequence from figure 5(a) to figure 5(i). Figure 5(a) gives the data and the longest period component, trend, which is the residue trend, not an IMF. By itself, the fitting of the trend is quite impressive, and it is very physical: the gradual decrease of the mean FID signal's amplitude indicates the lack of energy from the initially and the decrease of signal to noise ratio (SNR). As the mean amplitude decreases, the amplitude of the fluctuation increases. If we add the next longest period component, IMF9, the trend of the sum, IMF9 + trend, takes a remarkable turn, and the fitting to the data looks greatly improved as shown in figure 5(b). By successively adding more components with increasing frequency, we have the results in the series of figure 5(c) to figure 5(i). The gradual change from the monotonic trend to the final reconstruction is illustrative by itself. By the time we reach the sum of IMF components up to IMF3 in figure 5(g), we essentially have recovered all the energy containing eddies already. The components with the highest frequencies add little more energy, but they make the data look more complicated. In fact, the highest frequency component is probably not physical, for the digitizing rate of the ADC is too slow to capture the high-frequency variations. As a result, the data are jagged artificially by the digitalizing steps at this frequency.

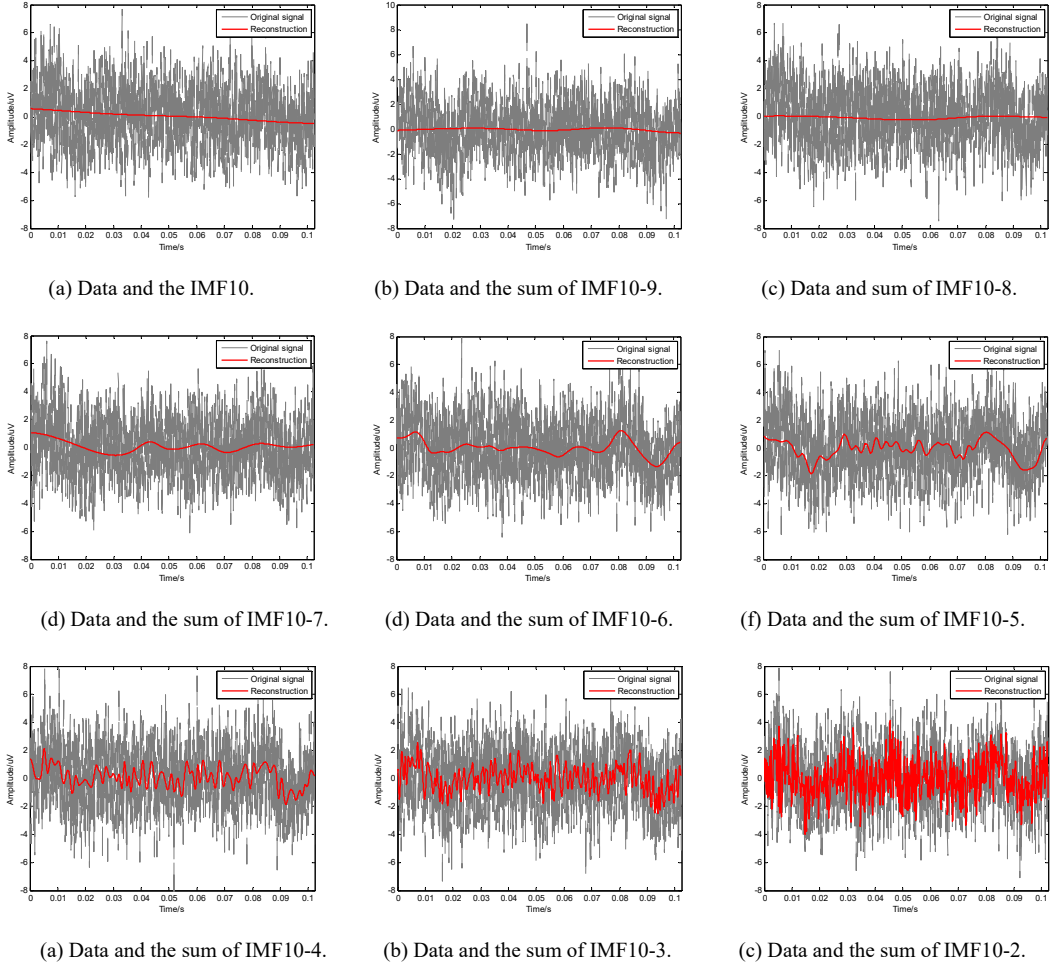

Fig. 5. Numerical proof of the completeness of the EMD through reconstruction of the original data from the IMF components.

## C. The Hilbert Spectrum

Having obtained the intrinsic mode function components, we will have no difficulties in applying the Hilbert transform to each component, and computing the instantaneous frequency according to equation 3.7. After preforming the Hilbert transform on each IMF component, we can express the data based on equation 3.9. Although the Hilbert transform can treat the monotonic trend as part of a longer oscillation, the energy involved in the residual trend the energy could be overpowering. In consideration of the uncertainty of longer trend, and in the interest of information contained in the other lower-energy and higher frequency components, the final non-IMF component should be left out [19]. It, however, could be included, if physical considerations justify its inclusion. Equation 3.9 gives both the amplitude and the frequency of each component as functions of time, and also enables to represent the amplitude and the instantaneous frequency as functions of time in a three-dimensional plot, in which the amplitude can be contoured on the frequency-time plane. This frequency-time distribution of the amplitude is designed as the Hilbert amplitude spectrum, or represent energy density, then the squared values of amplitude can be substituted to produce the Hilbert energy spectrum just as well.

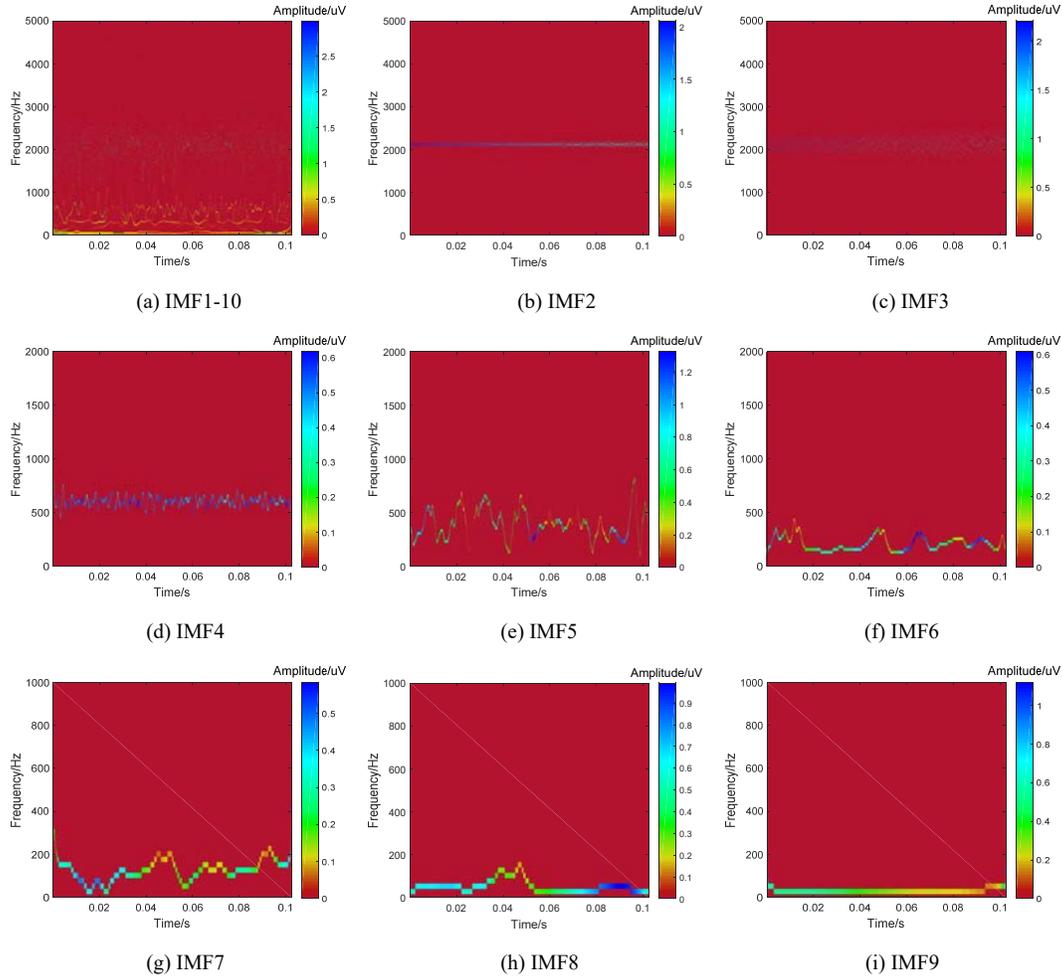

Fig. 6. The Hilbert spectrum for the FID signal data. The signal energy appears in skeleton lines representing each IMF.

The time-frequency distribution of all the IMFs are calculated using Hilbert transform, and the Hilbert spectrum in the color map format for the FID data is given in figure 6. Due to the IMF1 is more like a random signal having equal intensity at different frequencies and IMF10 is the trend of the original signal. In this case, IMF2-9 are more intuitive to be shown. We observe that the Hilbert spectrum appears only in the skeleton form with emphasis on the frequency variations of each IMF. If a more continuous form is preferred, the smoothing methods can be applied. In the smoothed form, the energy density and its trends of evolution as functions of frequency and time are easier to identify. In general, if more quantitative results are desired, the original skeleton presentation is better, as well as the smoothed presentation. From figure 6, we realize that, before adopted the EMD, the energy of FID signal distributed discretely, such as the white noise in the entire frequency band due to its equal intensity, the low frequency noise below 1000Hz etc. Therefore, the spectral characteristics of original or pure FID signal is not obvious, which has been drowned by noise. After processed by EMD, the main white noise is separated (refer to IMF1), the main energy of FID signal distributed around 2100Hz from IMF2. In addition, the other signals with different frequencies are also separated in each IMF by EMD.

*D. Contrast Test*

According to the experiments above, three noise reduction methods tests using the commonly methods: auto correlation and SVD, and the corresponding ensemble methods were conducted using the same platform. Table 1 and 2 show the comparing results. The ensemble means using HHT and EMD to process the signal first, then adopt the related filtering algorithm. To be specific, in table 1 and 2, A-ensemble means HHT + Auto correlation, while S-ensemble means HHT + SVD. In order to ensure the effectiveness, three sequences of actual FID signal were chosen randomly. We observe that all the SNRs are improved significantly after noise reduction by each of these methods, and the performance of ensemble are always better than the single method. Compared with only adopted auto correlation, the SNR approximately improved by 25% in average; while improved by 21% compared with implemented SVD only. In addition, when adopted auto correlation or SVD, the amplitude of original FID signal decreases a lot, while it almost unchanged when using the ensemble method. The original and valuable information of FID signal have been remained, as well as the less energy leak.

Table 1: Comparing results of different methods.

|  | Sequence | Unprocessed | Auto correlation | A-ensemble | SVD | S-ensemble |
| --- | --- | --- | --- | --- | --- | --- |
| Amplitude (uV) | First | 1.55 | 1.42 | **1.54** | 1.44 | **1.54** |
|  | Second | 1.62 | 1.49 | **1.59** | 1.53 | **1.60** |
|  | Third | 1.71 | 1.61 | **1.70** | 1.64 | **1.69** |
| SNR (dB) | First | 1.85 | 6.21 | **7.32** | 7.58 | **8.68** |
|  | Second | 1.94 | 6.35 | **7.44** | 7.59 | **8.85** |
|  | Third | 2.13 | 6.58 | **7.62** | 7.63 | **9.01** |

Table 2: Growth rate of SNR by different methods.

|  | Sequence | Auto correlation | A-ensemble | SVD | S-ensemble |
| --- | --- | --- | --- | --- | --- |
| Growth rate | First | 2.35 | **2.96** | 3.10 | **3.64** |
|  | Second | 2.27 | **2.84** | 2.91 | **3.56** |
|  | Third | 2.09 | **2.58** | 2.58 | **3.23** |
| Average |  | 2.24 | **2.79** | 2.86 | **3.48** |

From the results mentioned above, the proposed signal processing method, Hilbert-Huang transform can effectively realize the components visualization of FID signal, and the Hilbert spectrum further show each of the IMF component of FID signal more clearly. By the way, each IMF is obtained from the decomposition with this characteristic, and instantaneous frequency of each can be computed. Finally, by analyzing the resulting instantaneous frequencies, in figure 4 for practical FID signal decomposition, was identified a relatively pure FID signal in IMF2 (in this case), those are related to the noise suppression for FID signal. Moreover, it can also be adopted to prepossess the FID signal before using some kinds of filtering algorithm for tuning.

## V. Discussion

Generally, EMD, Fourier, and wavelets are all used to decompose signals, however, EMD is fundamentally different from the other two. The HHT is always regarded as a generalized Fourier transform in the sense that the decomposition of the signal of interest by HHT, leads to both amplitude and frequency varying time signals. A comparative summary of Fourier, wavelet and HHT analysis is given in table 3.

Table 3: Comparing results of different methods.

|  | Fourier | Wavelet | HHT |
| --- | --- | --- | --- |
| Basis | Prior | Prior | Adaptive |
| Presentation | Energy-frequency | Energy-time-frequency | Energy-time-frequency |
| Frequency | Global uncertain | Region uncertain | Local certainty |
| Nonlinear | × | × | √ |
| Non-stationary | × | √ | √ |
| Feature extraction | × | Continuous: √ Discrete: × | √ |

From table 3, we realize that the HHT is indeed a powerful method for analyzing data from nonlinear and non-stationary processes. It is based on an adaptive basis, and the frequency is derived from differentiation rather than convolution. Moreover, HHT is not limited by the uncertainly principle, and presents the results in time-frequency-energy space for feature extraction. Although HHT and wavelets provide complementary results, HHT is particularly attractive when analyzing signals from complex systems since the EMD decomposes the signal into a finite and small number of IMFs, which represent different scales of the original time series and physically meaningful modes. Whereas, choosing the wrong basis function can greatly increase the number of terms required to fit the time series.

## VI. Conclusion

With the measurement results of newly designed FID signal noise measurement platform, the noise components' character of FID signal is carefully studied in this paper. Based on the method of Hilbert-Huang for treating FID signal, new ways of processing resulted, such as feature extraction, filtering and noise elimination methods. HHT method can remain more valuable detail of the signal, because it can not only prevent the energy leak, but also help to find intrinsic features if the FID signal for short time analysis. Is possible to reduce time computing in FID signal just to further eliminate the noise components. Is possible to avoid some traditional calculations, only identifying the main frequencies involved to a weak magnetic field fluctuation. Moreover, before using relevant methods to process FID signal in any applications, the HHT method can be adopted in the preprocessed step, which could effectively reduce the noise influence on outdoor, and improve the proton precession magnetometer's anti-interference capability and system stability and reliability.

## Knowledge

This work is partly supported by the National Natural Science Foundation of China (Grant No. 41474158, 41504137), the National Key Scientific Instrument and Equipment Development Project of China (Grant No. 2014YQ100817) and the fund of Science and Technology on Near-Surface Detection Laboratory (Grant No. TCGZ2015A008).